\documentstyle[12pt]{article}
\def\Journal#1#2#3#4{{#1} {\bf #2},#3, (#4)}
\def\NPA{{\em Nucl.Phys.} A}

\def\PRL{\em Phys.Rev.Lett.}
\def\PRC{{\em Phys.Rev.} C}

\def\PLB{{\em Phys.Lett.} B}

\def\JP{{\em J.Phys.} G}
\begin{document}
\begin{titlepage}


\vspace{1cm}

\centerline{\large \bf POSSIBLE DIBARYONS WITH STRANGNESS S=-5
\footnote{this work was partly supported by the National Natural Science
Foundation of China}}

\vspace{1cm}

\centerline{Q.B.Li${^a}$, P.N.Shen$^{b,a,c,d}$}

\vspace{1cm}

{\small
{
\flushleft{\bf  $~~~$a.Institute of High Energy Physics,Chinese Academy
of Science,}
\vspace{-0.2cm}
\flushleft{\bf  $~~~~~$P.O.Box 918(4), Beijing  100039, China}
\flushleft{\bf  $~~~$b.Chinese Center of Advanced Science and Technology,
(World }
\vspace{-0.2cm}
\flushleft{\bf  $~~~~~$Laboratory), P.O.Box 8730, Beijing 100080, China}
\flushleft{\bf  $~~~$c.Institute of Theoretical Physics,Chinese Acedemy
of Sciences,}
\vspace{-0.2cm}
\flushleft{\bf  $~~~~~$P.O.Box 2735, Beijing 100080, China}
\flushleft{\bf  $~~~d.$ Center of Theoretical Nuclear Physic, National Lab
of Heavy Ion}
\vspace{-0.2cm}
\flushleft{\bf  $~~~~~$Accelerator, Lanzhou 730000, China}
}}

\vspace{1cm}

\centerline{\bf Abstract}

\baselineskip 30pt
\noindent
      In the framework of $RGM$, the binding energy of the six quark system 
with strangeness s=-5 is systematically investigated under the $SU(3)$ 
chiral constituent quark model. The single $\Xi^*\Omega$ channel calculation
with spins S=0 and 3 and the coupled $\Xi\Omega$ and $\Xi^*\Omega$ channel
calculation with spins S=1 and 2 are considered, respectively. The results 
show following observations: In the spin=0 case, $\Xi^* \Omega$ is a bound 
dibaryon with the binding energy being $80.0 \sim 92.4 MeV$. 
In the S=1  case, $\Xi\Omega$ is also a bound dibaryon. Its binding energy is 
ranged  from $26.2 MeV$ to $32.9 MeV$. In the S=2 and S=3 cases, no evidence of 
bound dibaryons are found. The phase shifts and scattering
lengths in the S=0 and S=1 cases are also given.
\end{titlepage}

\baselineskip 30pt

\vspace{0.3cm}
       Since Jaffe predicted the H particle in 1977 \cite{1}, dibaryon has been 
an important object to be investigated both theoretically and experimentally. 
Because this object is supposed to be a color singlet multi-quark system within 
a sufficiently smaller volume, the quark-gluon degrees of freedom become dominant. 
No doubt, studying this object can enrich our knowledge of the strong 
interaction in the short-range and can further prove
and complete the basic theory of strong interaction, Quantum Chromodynamics
$(QCD)$. In past twenty years, many dibaryons were proposed.
Among them, some are strangenessless such as $d^*$ \cite{2,3},
$d'$ \cite{4} and etc., and the others carry strangeness \cite{2,5}.
Moreover, as is well known, introducing strangeness and even heavier
flavor opens a new area to study strong interaction. It enables us to
further understand and deal with the effect of the
non-perturbative $QCD$ $(NPQCD)$ and to 
refine our knowledge of the strong interaction and, consequently, the 
hadronic structure by employing new model theories in the quark-gluon 
degrees of freedom via a variety of the new experimental 
data. Investigating
dibaryon with strangeness is just one of the most interesting subject in this 
aspect. Recently, Yu et al reported that in a system with higher strangeness,
it would be  highly possible to find bound dibaryons \cite{5}. They predicted the
existence of the $\Omega\Omega$ (S=0, T=0, L=0) 
and $\Xi\Omega$ (S=1, T=1/2, L=0) dibaryons, 
where S, T and L denote the spin, isospin and angular momentum, 
respectively. In this letter, we would systematically study the possible existence of bound dibaryons in the system with strangeness $s=-5$.

\vspace{0.3cm}

As is well known, the basic theory for studying dibaryon should be $QCD$ theory.
However, because of the complexity of $NPQCD$ effect at the lower energy region,
for practice, one has to develop $QCD$-inspired models. Among those models, the SU(3)
chiral-quark model is one of the most successful ones \cite{6}.
By employing this model, one can explain not only the single baryon properties 
\cite{7} but also the scattering data of the $N-N$ and $Y-N$ processes \cite{6}.
Moreover, the resultant  binding energy of H \cite{8} is consistent with the
experimental data available \cite{9,10}. 
Thus, it is reasonable to study the bound state
problem in the case with s=-5 by using this model.

\vspace{0.3cm}

In the model, the potential between the $i-$th and $j-$th constituent 
quarks can be written as
\begin{eqnarray}
       V_{ij}~=~\sum_{i<j}(V^{conf}_{ij}~+~V^{ch}_{ij}
            ~+~V^{OGE}_{ij})  .
\end{eqnarray}
In Eq.(1), the confinement potential $V^{conf}_{i,j}$ describes the long range 
effect of $NPQCD$, the one-gluon exchange potential $(OGE)$ basically depicts 
the short-range perturbation $QCD$ $(PQCD)$ effect. The potential induced
by the chiral-quark-field coupling is in the form of 
\begin{eqnarray}
       V^{ch}_{ij}~=~\sum_{a}V_{\pi_a}({\bf r}_{ij})
                ~+~\sum_{a}V_{\sigma_a}({\bf r}_{ij}) 
\end{eqnarray}
and mainly signifies the medium-range $NPQCD$ effect. In this expression, the
subscripts $\pi_a$ and $\sigma_a$ represent pseudoscalar mesons $\pi$, $k$, 
$\eta$ and $\eta'$ and scalar mesons $\sigma$, $\sigma'$, $\kappa$ and $\epsilon$, respectively. The explicit forms of these potentials can be found in Ref.\cite{6}.

\vspace{0.3cm}

The wave function of the single baryon can be expressed as

\begin{eqnarray}
\setlength{\unitlength}{0.33cm}
\begin{picture}(40,11)(4,0)
\multiput(6,9)(0,1){3}{\framebox(1,1)}
\put(11,10){$=$}
\multiput(12.5,10)(1,0){3}{\framebox(1,1)}
\put(18.5,10){$\times$}
\multiput(20.5,9)(0,1){3}{\framebox(1,1)}
\put(7,8){$OFSC$}
\put(15.5,9){$OFS$}
\put(22,8){$c$}
\put(23,10){$For~deculplet~baryons~(S=3/2)$}
\put(11,5){$=$}
\put(12,5){$\frac{1}{\sqrt{2}}$}
\put(14,5){${\bf(}$}
\multiput(15,5)(1,0){2}{\framebox(1,1)}
\put(15,4){\framebox(1,1)}
\put(17,6){$MS$}
\put(17,3){$OF$}
\put(19,5){$\times$}
\multiput(20,5)(1,0){2}{\framebox(1,1)}
\put(20,4){\framebox(1,1)}
\put(22,6){$MS$}
\put(22,3){$S$}
\put(24,5){$+$}
\multiput(25,5)(1,0){2}{\framebox(1,1)}
\put(25,4){\framebox(1,1)}
\put(27,6){$MA$}
\put(27,3){$OF$}
\put(29,5){$\times$}
\multiput(30,5)(1,0){2}{\framebox(1,1)}
\put(30,4){\framebox(1,1)}
\put(32,6){$MA$}
\put(32,3){$S$}
\put(34,5){${\bf)}$}
\put(35,5){$\times$}
\multiput(36,3)(0,1){3}{\framebox(1,1)}
\put(37,2){$c$}
\put(23,0){$For~octet~baryons~(S=1/2).$}
\end{picture}
\end{eqnarray}
Due to the flavor symmetry breaking, the wavefunctions in the orbit and flavor 
spaces are always associated. The orbital wave function of the $i-$th quark 
($i$ can be either up, down or strange quark) can be written as
\begin{eqnarray}
    \Phi_i({\bf r}_i)~=~(1/{\pi}b^{2}_{i})^{3/4}exp[-({\bf r}_i-{\bf R})^2/2b^{2}_{i}] ,
\end{eqnarray}
where $\bf R$ is the coordinate vector of the center of mass motion of the
baryon. The width parameter $b_i$ is associated with the oscillator frequency
$\omega$ by the constituent quark mass $m_i$
\begin{eqnarray}
   \frac{1}{b^{2}_{i}}~=~m_i\omega.
\end{eqnarray}
Then the wave function of the dibaryon can be written in the framework
of the Resonating Group Method $(RGM)$ as
\begin{eqnarray}
  {\Psi}_{6q}~=~{\cal A}[{\Phi}_A{\Phi}_B\chi({\bf R}_{AB})Z({\bf R}_{CM})],
\end{eqnarray}
where ${\chi}({\bf R_{AB}})$ is the trial relative wave function between 
clusters A and B, $Z({\bf R}_{CM})$ represents the CM  wave function of
the six quark system and ${\cal A}$ denotes the antisymmetrizer.
Expanding unknown ${\chi}({\bf R_{AB}})$ by employing well-defined
basis wavefunctions, such as Gaussian functions, one can 
solve RGM bound state equation to obtain
eigenvalues and  corresponding wave functions,
simultaneously. The details  of solving RGM bound-state problem can be
found in Refs. \cite{11,12}.

\vspace{0.3cm}

All the model parameters employed in this letter are those used in our previous 
papers \cite{6,7,8,5,10}. These parameters can be determined 
by the mass splittings among $N$, $\Delta$, $\Lambda$, $\Sigma$ and $\Xi$,
respectively, and the stability conditions of the octet $(S=1/2)$ and
decuplet $(S=3/2)$ baryons, respectively. The values of the parameters
are tabulated in Table 1.

\vspace{0.3cm}

\centerline{\bf {Table 1~~~~Model parameters$^{\dag}$}}
\begin{center}
\begin{tabular}{ccc|ccc}\hline
                              & $~~~~~$    Set1  & $~~~~~$  Set2 &                                   & $~~~~~$  Set1  & $~~~~~$  Set2    \\\hline
  $m_u~(MeV)$                 & $~~~~~$    313   & $~~~~~$  313  &                                   &                &                  \\
  $m_s~(MeV)$                 & $~~~~~$    470   & $~~~~~$  430  &                                   &                &                  \\
  $b_u~(fm)$                  & $~~~~~$    0.505 & $~~~~~$  0.53 &                                   &                &                  \\
  $m_\pi~(fm^{-1})$           & $~~~~~$    0.7   & $~~~~~$  0.7  & $\Lambda_\pi~(fm^{-1})$           & $~~~~~$  4.2   & $~~~~~$  4.2     \\
  $m_k~(fm^{-1})$             & $~~~~~$    2.51  & $~~~~~$  2.51 & $\Lambda_k~(fm^{-1})$             & $~~~~~$  4.2   & $~~~~~$  4.2     \\
  $m_\eta~(fm^{-1})$          & $~~~~~$    2.78  & $~~~~~$  2.78 & $\Lambda_\eta~(fm^{-1})$          & $~~~~~$  5.0   & $~~~~~$  5.0     \\
  $m_{\eta'}~(fm^{-1})$       & $~~~~~$    4.85  & $~~~~~$  4.85 & $\Lambda_{\eta'}~(fm^{-1})$       & $~~~~~$  5.0   & $~~~~~$  5.0     \\
  $m_\sigma~(fm^{-1})$        & $~~~~~$    3.17  & $~~~~~$  3.04 & $\Lambda_\sigma~(fm^{-1})$        & $~~~~~$  4.2   & $~~~~~$  7.0     \\
  $m_{\sigma'}~(fm^{-1})$     & $~~~~~$    4.85  & $~~~~~$  4.85 & $\Lambda_{\sigma'}~(fm^{-1})$     & $~~~~~$  5.0   & $~~~~~$  5.0     \\
  $m_\kappa~(fm^{-1})$        & $~~~~~$    4.85  & $~~~~~$  4.85 & $\Lambda_\kappa~(fm^{-1})$         & $~~~~~$  5.0   & $~~~~~$  7.0     \\
  $m_\epsilon~(fm^{-1})$      & $~~~~~$    4.85  & $~~~~~$  4.85 & $\Lambda_\epsilon~(fm^{-1})$      & $~~~~~$  5.0   & $~~~~~$  7.0     \\
  $g_u$                       & $~~~~~$  0.936   & $~~~~~$ 1.010 &                                   &                &                  \\
  $g_s$                       & $~~~~~$  0.924   & $~~~~~$ 0.965 &                                   &                &                  \\
  $a_{uu}~(MeV/fm^2)$         & $~~~~~$  54.34   & $~~~~~$ 34.37 & $a^0_{uu}~(MeV)$                  & $~~~~~$ -47.69 & $~~~~~$ -19.76   \\
  $a_{us}~(MeV/fm^2)$         & $~~~~~$  65.75   & $~~~~~$ 39.59 & $a^0_{us}~(MeV)$                  & $~~~~~$ -41.73 & $~~~~~$ -7.73    \\
  $a_{ss}~(MeV/fm^2)$         & $~~~~~$  102.97  & $~~~~~$ 69.91 & $a^0_{ss}~(MeV)$                  & $~~~~~$ -45.04 & $~~~~~$ -11.54   \\\hline               &
\end{tabular}
\flushleft{$\dag$ {\footnotesize {By employing either Set 1 or Set 2, 
the experimental $NN$ and $NY$ scattering data as well as}}}

\vspace{-0.9cm}

\flushleft{$~~$ {\footnotesize {some properties of single 
baryons can be well reproduced.}}}
\end{center}

\vspace{0.3cm}

The six-quark system with strangeness $s=-5$, isospin $T=1/2$ and spin $S=0$
is studied first. This is a single channel calculation. The model parameters
used in the first step is Set 1 in Table 1, which is frequently used in our 
previous investigations \cite{6,7,8}. The resultant binding energy
($B_{\Xi^{*}\Omega}$) and corresponding root-mean-square radius (RMS)
are tabulated in Table 2, respectively. It is shown that the binding energy is
$92.4 MeV$ and the corresponding RMS is $0.71 fm$. It indicates that this is a 
bound dibaryon. However, $\Xi^*$ is not a stable particle and
easily subjects the strong decay $\Xi^*~\rightarrow~\Xi~+~\pi$.
To detect $\Xi^*\Omega$ dibaryon in the experiment easily, it is better to have
the mass of $\Xi^*\Omega$ dibaryon lower than the threshold of the
$\Xi~+~\Omega~+~\pi$ channel, or the binding energy of $\Xi^* \Omega$
$$B_{\Xi^*\Omega}~>~-(M_\Xi~+~M_\pi~-~M_{\Xi^*})~=~76 MeV.$$
In fact, the predicted mass of $\Xi^*\Omega$ dibaryon is about $16.4 MeV$ below 
the $\Xi\Omega\pi$ threshold, namely, this dibaryon is stable against the 
strong decay $\Xi^*~\rightarrow~\Xi~+~\pi$.

\vspace{0.3cm}

The model parameter-dependence of the binding energy of $\Xi^*\Omega$ 
is then examined. The calculated result shows that the binding energy increases
if the masses of the s quark and $\kappa$ meson and the cut-off masses of the $\sigma$,
$\sigma'$ and $\epsilon$ mesons increase and the masses of the $\sigma$,
$\sigma'$ and $\epsilon$ mesons and the cut-off mass of the $\kappa$
meson decrease. In particular, in a six-quark system with higher strangeness
number, increasing $b_u$ would make the binding energy smaller. 
Therefore, another
set of parameters in limits, Set 2, which can fit all mentioned empirical 
data, is also employed to estimate the binding energy. The results are  
shown in Table 2.
The lower limit of the binding energy of $\Xi^*\Omega$ is $80.0 MeV$ and the
corresponding RMS is $0.76 fm$. Further considering the strong
decay of $\Xi^*$ into $\Xi\pi$, the mass of $\Xi^*\Omega$ is $4.0 MeV$ lower 
than the $\Xi\Omega\pi$ threshold. Anyway, the $\Xi^*\Omega$ dibaryon is a
stable particle against the strong decay $\Xi^*~\rightarrow~\Xi~+~\pi$, and 
its mass is 4.0 MeV-16.4 MeV below the $\Xi\Omega\pi$ threshold.

\centerline{\bf {Table 2~~~~Binding energy $B_{AB}$ and RMS of $\Xi^* \Omega (S=0)$
 and $\Xi\Omega (S=1)$}$^{\dag}$}

\begin{center}
\begin{tabular}{|c|c|c|c|c|c|c|}
\hline
$~$          & \multicolumn{2}{|c|}{one channel}
& \multicolumn{2}{|c|}{one channel}
& \multicolumn{2}{|c|}{couple channel} \\
 Channel  &  \multicolumn{2}{|c|}{$\Xi^*\Omega (S=0)$} 
& \multicolumn{2}{|c|}{$\Xi\Omega (S=1)$}
 & \multicolumn{2}{|c|}{$\Xi\Omega-\Xi^*\Omega (S=1)$}\\
\cline{2-7}
  & $B_{\Xi^*\Omega}$ & $RMS$ & $B_{\Xi\Omega}$ & $RMS$
  & $B_{\Xi\Omega}$ & $RMS$  \\
  &$(MeV)$ & $(fm)$ & $(MeV)$ & $(fm)$  & $(MeV)$ & $(fm)$  \\  \hline
 Set1 &  92.4 & 0.71  &  9.6 & 1.02 & 32.9 & 0.78\\  \hline
 Set2 &  80.0 & 0.76  &  6.3 & 1.12 & 26.2 & 0.85\\  \hline
\end{tabular}
\end{center}
$\dag$ {\footnotesize {$B_{AB}$ denotes the binding energy between
clusters A and B. }}

\vspace{0.3cm}

Next, the six-quark system with $s=-5$, $T=1/2$ and $S=1$ is studied.
$\Xi\Omega (S=1,~T=1/2,~L=0)$ was predicted in the single channel RGM 
calculation \cite{5}. As is well known, there exists another possible
channel $\Xi^*\Omega$ having same quantum numbers. Although the threshold 
of the $\Xi^*\Omega$ channel is about $200 MeV$ higher than that of the 
$\Xi\Omega$ channel, because the cross-channel interaction matrix element
might not be small, it is necessary to check whether the additional 
$\Xi^* \Omega$ channel exerts substantial
effect on the single channel prediction. In the calculation, the L=0 and L=2
states for each channel are considered. The results with Set 1 and Set 2 are 
collected in Table 2. From these data, one sees that the additional
$\Xi^* \Omega$ channel indeed gives sizable contribution and cannot be ignored. 
The resultant binding energy of $\Xi\Omega$ is ranged from $26.2 MeV$ to 
$32.9 MeV.$ and the corresponding $RMS$ of $\Xi\Omega$ is in the region of 
$0.85 \sim 0.78 fm$. Again, $\Xi\Omega$ is a stable dibaryon.

\vspace{0.3cm}

Finally, the six-quark system with $s=-5$, $T=1/2$ and $S=2$ and 3 are
studied. In the $S=2$ case, it is a coupled $\Xi\Omega-\Xi^* \Omega$
channel calculation and in the $S=3$ case, it is just a single $\Xi^* \Omega$
channel calculation. With either parameter Set 1 or Set 2, no evidence of
the bound dibaryon is found.

\vspace{0.3cm}

To provide more information about the dibaryons with strangeness $s=-5$
and crosscheck dibaryons' binding behaviors, we further demonstrate the S-wave phase
shifts of the $\Xi^*\Omega(S=0)$ and the $\Xi\Omega(S=1)$ with parameters Sets 1
and 2 in Figs. 1 and 2, respectively. 
In these figures, solid curves are the results with Set 1 and the dashed
curves denote the results with Set 2. According to these phase shifts,
one can easily estimate the scattering length $a$. The results are
presented in Table 3:

\centerline{\bf {Table 3~~~~The Scattering length $a$ }}
\begin{center}
\begin{tabular}{|c|c|c|}
\hline
        &   one channel          &     two channel\\
        &   $\Xi^*\Omega~(S=0)$  & $\Xi\Omega~-~\Xi^*\Omega~(S=1)$\\ \hline
  Set 1 &   $-1.18~(fm)$         &     $-1.55~(fm)$\\ \hline
  Set 2 &   $-1.31~(fm)$         &     $-1.73~(fm)$\\ \hline
\end{tabular}
\end{center}
Both phase shifts and scattering lengths are consistent with our
above findings.

\vspace{0.3cm}

  As the conclusion, we announce that in the $s=-5$ sector of the six-quark
system, there may exist two bound states or bound dibaryons. One of them is
$\Xi^*\Omega$ with $S=0$, $T=1/2$ and $L=0$. The binding energy, $RMS$
and the corresponding scattering length of this system are ranged from
$80.0 MeV$ to $92.4 MeV$, from $0.76 fm$ to $0.71 fm$ and from $-1.18 fm$
to $-1.31 fm$, respectively. Because the mass of the $\Xi^*\Omega$
particle is $4.0 MeV \sim 16.4 MeV$ below the $\Xi\Omega\pi$ threshold, this
particle should be a stable dibaryon against the strong decay
$\Xi^*~\rightarrow~\Xi~+~\pi$, but it  still 
can weakly decay into $\Lambda\Omega\pi\pi$, $\Xi\Lambda{k}\pi$, etc. The width of $\Xi^*\Omega$ can roughly be estimated in the following. It might be 
narrower than that of single $\Xi^*$ which decays through strong 
mode $\Xi^*\rightarrow\Xi\pi$ and comparable to that of $\Xi$ or $\Omega$ 
which decay through weak modes $\Xi\rightarrow\Lambda\pi$, $\Omega\rightarrow
\Lambda{k}$, etc. Another possible stable dibaryon is 
$\Xi\Omega$ $(S=1,T=1/2,L=0)$. In the coupled $\Xi\Omega-\Xi^* \Omega$
(with $L=0$ and 2) channel approximation,
its binding energy, $RMS$ and the corresponding scattering length are in 
the regions of  $26.2 \sim 32.9 MeV$, $0.85 \sim 0.78 fm$ and 
$-1.73 \sim -1.55 fm$, respectively. Since $\Xi^*$, $\Xi$ and $\Omega$ are
secondary particles, we suggest to search $\Xi^* \Omega~(S=0,~T=1/2)$ and
$\Xi\Omega~(S=1,~T=1/2)$ dibaryons in the heavy ion collision.


\newpage

\begin{thebibliography}{99}
\bibitem{1}
R.J.Jaffe, \Journal{\PRL}{38}{195}{1977}.
\bibitem{2}
K.Yazaki,{\em Prog.Theor.Phys.Suppl}.91,146(1987);\\
P.J.G.Mulders, A.T.Aerts, J.J.Swarts, \Journal{\PRL}{40}{1543}{1978};\\
F.Wang, C.H.Wu, L.J.Teng, T.Goldman, \Journal{\PRL}{69}{2901}{1992};\\
F.Wang, J.L.Ping, C.H.Wu, L.J.Teng, T.Goldman, \Journal{\PRC}{51}{3411}{1955};\\
X.Q.Yuan, Z.Y.Zhang, Y.W.Yu, P.N.Shen, submitted to {\em Phys.Rev.C}.
\bibitem{3}
Y.Kamaen, T.Fujita, \Journal{\PRL}{38}{471}{1977}.
\bibitem{4}
R.Bilger,{\it et.al.},  \Journal{\PLB}{269}{247}{1991};\\
R.Bilger, H.A.Clement, \Journal{\PRL}{71}{42}{1993};\\
G.Wagner, L.Ya.Glozman, A.J.Buchmann, A.Faessler, \Journal{\NPA}{594}{263}{1995};\\
A.J.Buchmann, G.Wagner, A.Faessler, \Journal{\PRC}{57}{3340}{1998}.
\bibitem{5}
Y.W.Yu, Z.Y.Zhang, X.Q.Yuan,{\em Commun.Theor.Phys}.31(1998);\\
Y.W.Yu,Z.Y.Zhang,X.Q.Yuan,{\em High Ener.Phys.Nucl.Phys}.(in press).
\bibitem{6}
Z.Y.Zhang, Y.W.Yu, P.N.Shen, L.R.Dai, A.Faessler, U.Straub,
\Journal{\NPA}{625}{59}{1997};\\
S.Yang, P.N.Shen, Z.Y.Zhang, Y.W.Yu,  \Journal{\NPA}{635}{146}{1998}.
\bibitem{7}
P.N.Shen, Y.B.Dong, Y.W.Yu, Z.Y.Zhang, T.S.H.Lee, \Journal{\PRC}{55}{204}{1997};\\
H.Chen, Z.Y.Zhang, {\em High Ener.Phys.Nucl.Phys},1997.
\bibitem{8}
P.N.Shen, Z.Y.Zhang, Y.W.Yu, X.Q.Yuan, S.Yang, {\em BIHEP-TH-1997-50}; \Journal{\JP}{25}{}{1999} (in press).
\bibitem{9}
U.Straub, Z.Y.Zhang {\it et al.}, \Journal{\PLB}{200}{241}{1988}.
\bibitem{10}
P.N.Shen, Z.Y.Zhang, X.Q.Yuan, S.Yang, (in press).
\bibitem{11}
"A unified theory of the nucleus", K.Wildermuth, Y.C.Tang, Academic Press.Inc.
\bibitem{12}
U.Straub, Z.Y.Zhang {\it et al.}, \Journal{\NPA}{483}{686}{1988}
\end{thebibliography}
\end{document}